\documentclass[american,twocolumn,prl]{revtex4-2}
\usepackage[T1]{fontenc}
\usepackage[latin9]{inputenc}
\usepackage{array}
\usepackage{amstext}
\usepackage{amssymb}
\usepackage{graphicx}

\makeatletter

\providecommand{\tabularnewline}{\\}

\makeatother

\usepackage{babel}
\begin{document}
\title{Internally Driven $\beta$-plane Plasma Turbulence Using the Hasegawa-Wakatani System}
\author{{\"O}. D. G\"urcan}
\affiliation{Laboratoire de Physique des Plasmas, CNRS, Ecole Polytechnique, Sorbonne
Universit\'e, Universit\'e Paris-Saclay, Observatoire de Paris,
F-91120 Palaiseau, France}
\begin{abstract}
General problem of plasma turbulence can be formulated as advection
of potential vorticity (PV), which handles flow self-organization,
coupled to a number of other fields, whose gradients provide free
energy sources. Therefore, focusing on PV evolution separates the
underlying linear instability from the flow self-organization, and
clarifies key spatial scales in terms of balances between various
time scales. Considering the Hasegawa-Wakatani model as a minimal,
nontrivial model of plasma turbulence where the energy is injected
internally by a linear instability, we find that the critical wavenumber
$k_{c}=C/\kappa$ where $C$ is the adiabaticity parameter and $\kappa$
is the normalized density gradient separates the adiabatic (or highly
zonostrophic) behavior for large scales from the hydrodynamic behavior
at small scales. In the adiabatic range the non-zonal part of the
wave-number spectrum goes from $E\left(k\right)\propto\gamma_{k}U^{-1}k^{-2}$
around the peak to $E\left(k\right)\propto\omega_{k}^{2}k^{-3}$ in
the ``inertial'' range, where $\gamma_{k}$ and $\omega_{k}$ are
the linear growth and frequency and $U$ is the rms zonal velocity.
This proposed spectrum fits very well for the large $k_{c}$ case,
where the bulk of the spectrum is in the adiabatic range. In contrast
for small $k_{c}$, we get the usual forward enstrophy cascade with
$E\left(k\right)\propto\epsilon_{W}^{2/3}k^{-3}$, where $\epsilon_{W}$
is the enstrophy dissipation. In contrast for $k_{c}\approx1$, the
system transitions to hydrodynamic forward enstrophy cascade right
after the injection range, with zonal flows at large scales and forward
enstrophy cascade at small scales. Note that $k_{c}$, can also be
used as a proxy for the scale at which the system switches from wave-dominated
(i.e. $E\left(k\right)\propto\omega_{k}^{2}k^{-3}$) to hydrodynamic
(i.e. $E\left(k\right)\propto\epsilon_{W}^{2/3}k^{-3}$) spectra usually
denoted by $k_{\beta}$ in geophysical fluid dynamics. It is argued
that the ratio $R_{\beta}\equiv k_{\beta}/k_{\text{peak}}\approx k_{c}/k_{\text{peak}}$
where $k_{\text{peak}}$ is the peak wave-number can be defined as
the zonostrophy parameter, and that the abundance of zonal flows vs.
eddies in near and far from ``marginality'' that is commonly formulated
in terms of the Kubo number in plasma problems can also be understood
in terms of the zonostrophy parameter, since $R_{\beta}$ increases
as we approach marginality.
\end{abstract}
\maketitle

\section{Introduction}

Understanding plasma turbulence and its self-regulation through generation
of large scale flows is an important challenge in predicting and controlling
turbulent transport in fusion devices. Micro-instabilities, driven
by background gradients, inevitably present in a magnetic confinement
device, generate collective, fluctuating electric fields that can
transport particles, heat and momentum towards the walls. These fluctuations
also generate large scale flows, called zonal flows through Reynolds
stresses\citep{diamond:05}. Zonal flows can then suppress the turbulence
that drives them through a mechanism of shear suppression\citep{biglari:90},
thus allowing the turbulence to self-regulate. This mechanism of flow
self-organization is well known, and generally understood in terms
of the nonlinear dynamics of potential vorticity\citep{gurcan:15},
and is closely related to the similar mechanism of layer formation
in geophysical fluid dynamics (GFD) \citep{vallis:93,vallis:book}.

On the other hand, fluctuations in plasma turbulence, especially in
scales where the instability mechanism is active are known to be roughly
quasi-linear, where the phases between fluctuations are dictated by
linear relations\citep{casati:09b}. This justifies our approach of
separating the instability mechanism, which is almost linear, from
flow self-organization through the PV equation, which is completely
nonlinear. Note that a generalization of the QLT is also commonly
used in the GFD community\citep{marston:16}, except that while plasma
QLT works beter when there are no zonal flows, the GFD type gQLT works
best when the zonal flows are the strongest. The current letter, in
addition to providing theoretical estimates of the wave-number spectra
in different limits, intends to bridge this gap between the two communities. 

\paragraph*{$\beta$-plane model of plasma turbulence:}

Consider an equation for the potential vorticity (PV) $\zeta$, which
can be written in general as:
\begin{equation}
\partial_{t}\zeta+\hat{\mathbf{z}}\times\nabla\Phi\cdot\nabla\zeta+\kappa\partial_{y}\widetilde{\Phi}=-\mu\zeta+\nu\nabla^{2}\zeta\;\text{,}\label{eq:pv}
\end{equation}
where $\kappa$ is its normalized background gradient $\Phi=\overline{\Phi}+\widetilde{\Phi}$
where $\overline{\Phi}$ is the zonally averaged part of the properly
normalized electrostatic potential playing the role of the stream
function, $\mu$ and $\nu$ are large scale friction and small scale
dissipation respectively, which serve to regularize the equation and
could also be replaced by hypo and hyper viscosities. Note that if
one considers a self-consistent, internally driven model such as the
Hasegawa-Wakatani system\citep{hasegawa:83,camargo:95} (with the
standard modified response), a large scale friction is not necessary
for steady state, and we can set $\mu=0$, however if we only consider
a linear dispersion relation in order to close (\ref{eq:pv}), the
inverse cascade becomes an issue and one has to include large scale
friction and hypodiffusion.

The definition of PV in terms of various fields of the system, and
its inversion (i.e. writing $\Phi$ in (\ref{eq:pv}) in terms of
$\zeta$ , or writing $\zeta$ in terms of $\Phi$ alone so that we
have a single field) gives us the details of the physics problem.
For example for dissipative drift waves $\zeta=n-\nabla^{2}\Phi$,
where $n$ is the electron density and the inversion requires that
we solve the continuity equation for density along with Eqn. (\ref{eq:pv}).
For ion temperature gradient driven (ITG) turbulence, pressure $P$
also enters the definition of PV through $\zeta=n-\nabla^{2}\Phi-P/\Gamma$
where $\Gamma$ is the specific heat ratio, and we need to solve an
equation for pressure as well, which is in turn coupled to a parallel
velocity equation\citep{horton:81}. Even the gyrokinetic equation
can be defined in terms of a potential vorticity conservation through
the gyrocenter density\citep{mcdevitt:10}, or the zeroth moment of
the gyrokinetic distribution function, where the inversion problem
becomes the solution of all the other moments.
\begin{figure*}
\begin{centering}
\includegraphics[width=0.99\textwidth]{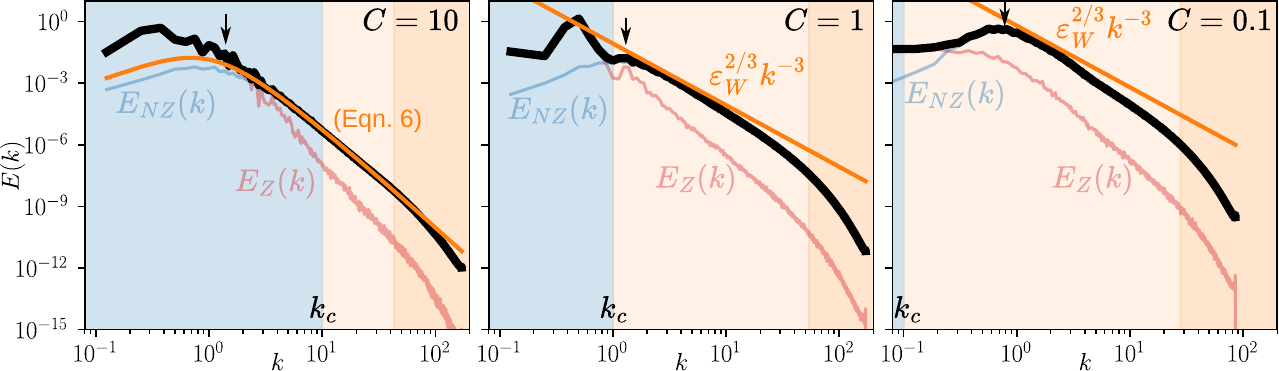}
\par\end{centering}
\caption{\label{fig:spectra}$E\left(k\right)$ in Hasegawa-Wakatani turbulence
for $C=10$ to $1$ to $0.1$ from left to right with $\kappa=1$,
so that the transition scale $k_{c}$ is varied by two orders of magnitude.
Here the blue region is adiabatic, light orange is hydrodynamic and
dark orange is dissipative regions, and the wavenumbers for which
$\gamma_{k}$ is maximum are shown with little black arrows. The solid
black curve is the kinetic energy spectra $E\left(k\right)$ computed
as $E\left(k\right)=\sum_{k\in k\pm\Delta k}k^{2}\Phi_{k_{x},k_{y}}^{2}$,
averaged over time in the saturated phase of simulations up to $t=5000$,
with a padded resolution of $4096^{2}$. The light red curve corresponds
to the zonal spectra $E_{Z}\left(k_{x}\right)=\sum_{k_{x}\in k_{x}\pm\Delta k_{x}}k_{x}^{2}\Phi_{k_{x},0}^{2}$
whereas $E_{NZ}\left(k\right)=E\left(k\right)-E_{Z}\left(k\right)$.
Note that the solid orange curves corresponding to the theoretical
spectra (as given by Eqn. 5, or the Kraichnan-Kolmogorov forward enstrophy
cascade with $\varepsilon^{2/3}k^{-3}$) are not multiplied by any
additional factors, they naturally fall on the numerical spectra.}
\end{figure*}

It may be possible in some cases to write a linear inversion for the
other fields. For example, the Hasegawa-Wakatani system could be inverted
approximately using the linear relation $n_{k}\rightarrow\sigma_{k}\Phi_{k}$
with $\sigma_{k}\approx\frac{\kappa k_{y}+iC}{\omega_{k}+iC}$ so
that:
\begin{equation}
\widetilde{\Phi}_{k}=\left(\sigma_{k}+k^{2}\right)^{-1}\widetilde{\zeta}_{k}\;\text{,}\label{eq:inv}
\end{equation}
where $\omega_{k}$ is the unstable solution of the linear dispersion
relation, which can be written in the inviscid limit as $\omega_{k}^{2}k^{2}+iC\omega_{k}\left(1+k^{2}\right)=i\kappa k_{y}C$,
with $C$ the adiabaticity parameter. Note that substituting (\ref{eq:inv})
into (\ref{eq:pv}) and linearizing, we get back the linear dispersion
relation. However linear inversion is problematic in particular because
the resulting system lacks the key saturation mechanism \emph{involving
the backreaction of the flow on the linear growth}. This can be remedied
in part by using a quasi-linear inversion, or eventually, fully nonlinear
inversion, which requires solving the continuity equation:
\begin{equation}
\partial_{t}n+\hat{\mathbf{z}}\times\nabla\Phi\cdot\nabla n+\kappa\partial_{y}\widetilde{\Phi}=C\left(\widetilde{\Phi}-\widetilde{n}\right)+D\nabla^{2}n\label{eq:dens}
\end{equation}
making it necessary to consider the full Hasegawa-Wakatani system
{[}i.e. Eqns. (\ref{eq:pv}) and (\ref{eq:dens}){]}, even if we are
only interested in PV evolution since this is the only way we can
infer $\widetilde{\Phi}$ that appears in (\ref{eq:pv}). Note that
from the perspective of PV evolution, the energy injection in plasma
turbulence is around $k_{y}\approx1$ (or smaller) and displays characteristics
of forward cascade either through potential enstrophy cascade carried
by waves, or enstrophy cascade through eddies. The inverse energy
``cascade'' is usually not important, since the scale at which the
zonal flows form is usually slightly larger than the injection scale,
and they can turn off the underlying instability drive. However if
one uses only the linear dispersion relation (instead of the full
nonlinear problem), to invert PV, while PV can still flatten, it can
not turn of the underlying instability drive, and thus the system
becomes pathological, necessitating large scale friction, but the
behaviour of the system with linear inversion and large scale friction
is different from that of self-consistent saturation without or with
little friction.

The key spatial scale in plasma $\beta$ plane turbulence, is the
transition scale between adiabatic and hydrodynamic behavior. For
the case of the Hasegawa-Wakatani model, this is a linear scale defined
as $k_{c}=C/\kappa$. More generally one can define this scale as
the scale where $\sigma_{kr}\approx\sigma_{ki}$ , (i.e. basically
if we define $\sigma_{k}\approx1+i\delta_{k}$ this is the scale where
$\delta_{k}\approx1$). This suggests that for scales larger than
this scale, we have the adiabatic behavior (i.e. $n_{k}\approx\Phi_{k}$)
whereas for scales smaller than this scale, the hydrodynamic behavior
follows. In three dimensions, the argument can be extended by noting
that $C$ goes like $k_{\parallel}^{2}$ \citep{barnes:11}. Note
that $k_{c}$ is closely related to $k_{\beta}$, commonly used in
GFD as the scale at which the system switches from inverse cascade
to Rossby wave turbulence (recall that, there the energy injection
is at small scales). Since the behavior of the spectra are completely
different in the adiabatic vs. hydrodynamic regimes (See table \ref{tab:scales}
for a detailed classification), we characterize everything with respect
to $k_{c}$ first. 

Considering first $k_{c}\gg1$ (e.g. $C\gg1$ with $\kappa\approx1$),
with a general growth rate $\gamma_{k}$ which has the most unst<able
mode that has a finite $k_{y}=k_{y0}\approx O\left(1\right)$ and
vanishing $k_{x}$ (the usual case in plasma turbulence), we can discuss
some features of the nonlinear saturation. Since strong zonal flows
are formed in this state, we can define the usual time scale associated
with the zonal flows as $\tau_{Z}^{-1}=Uk$ where $U$ is the rms
zonal velocity, and a hybrid time scale of the form $\tau_{gz}^{-1}\approx\left(Uk\gamma_{k}\right)^{1/2}$.
\begin{table}
\begin{tabular}{>{\raggedright}p{0.28\columnwidth}>{\raggedright}p{0.35\columnwidth}>{\raggedright}p{0.37\columnwidth}}
$E\left(k\right)$ Spectrum & Time scale balance & range\tabularnewline[3pt]
\hline 
\noalign{\vskip3pt}
$U\gamma_{k}k^{-2}$ & $\sqrt{Uk\gamma_{k}}\sim\tau_{n\ell}^{-1}$ & $k\sim k_{\text{peak}}\ll k_{c}$ \tabularnewline[3pt]
\noalign{\vskip3pt}
$\omega_{k}^{2}k^{-3}$ & $\omega_{k}\sim\tau_{n\ell}^{-1}$ & $k_{\text{peak}}<k<k_{c}$\tabularnewline[3pt]
\noalign{\vskip3pt}
$\varepsilon^{2/3}k^{-3}$ & $\varepsilon^{1/3}\sim\tau_{n\ell}^{-1}$ & $k>k_{\text{peak}}\lesssim k_{c}$\tabularnewline[3pt]
\hline 
\noalign{\vskip3pt}
\end{tabular}

\caption{\label{tab:scales}Limiting forms of the wave-number spectra, corresponding
time scale balances {[}here $\tau_{n\ell}^{-1}\equiv E^{1/2}\left(k\right)k^{3/2}${]}
and the range of wave-numbers for which they are applicable.}
\end{table}

We argue that, as long as the peak region is such that $k_{\text{peak}}<k_{c}$
it is this hybrid scale which is balanced by the nonlinear time $\tau_{n\ell}^{-1}\propto E\left(k\right)^{1/2}k^{3/2}$
around the peak of the spectrum, yielding the quasi-linear spectrum:
\begin{equation}
E\left(k\right)\approx U\gamma_{k}k^{-2}\;\text{,}\label{eq:Epeak}
\end{equation}
valid for $\gamma_{k}>0$. Note that wile the rms velocity $U$ is
an emergent feature, we can estimate it from the momentum balance
between the waves and flows, using something like $U\approx\left[\omega_{k}/k_{y}\right]_{\text{max}}$.
For example for the standard form of $\omega_{k}\propto\kappa k_{y}/\left(1+k^{2}\right)$,
we get $U=\kappa$. However a more general form of the momentum balance,
which can be written as $U\approx\left[\omega_{k}/k_{y}\right]_{\text{max}}-\left(\partial_{x}\overline{n}\right)_{\text{rms}}+\left(\partial_{x}\overline{\Omega}\right)_{\text{rms}}$can
not be solved in terms of linear quantities alone.

For $k_{c}>0.1$, The linear growth peaks around $k_{y}\approx1$
and is fairly localized to wavenumbers around its peak and becomes
negative at large $k$ due to viscosity. If the wave-number where
$\gamma_{k}<0$ is smaller than $k_{c}$, we continue with adiabatic
wave-turbulence and this time, it is the linear frequency $\omega_{k}$
which balances the nonlinear time, giving a $k$-spectrum of the form:
\begin{equation}
E\left(k\right)\approx\omega_{k}^{2}k^{-3}\;\text{.}\label{eq:Ew}
\end{equation}
The two solutions can actually be put together by using $\gamma_{k}\rightarrow\gamma_{k}+\nu k^{2}$
(which corresponds to the inviscid part of $\gamma_{k}$, which asymptotes
to zero instead of becoming negative), so that we can write:
\begin{figure*}
\begin{centering}
\includegraphics[width=0.99\textwidth]{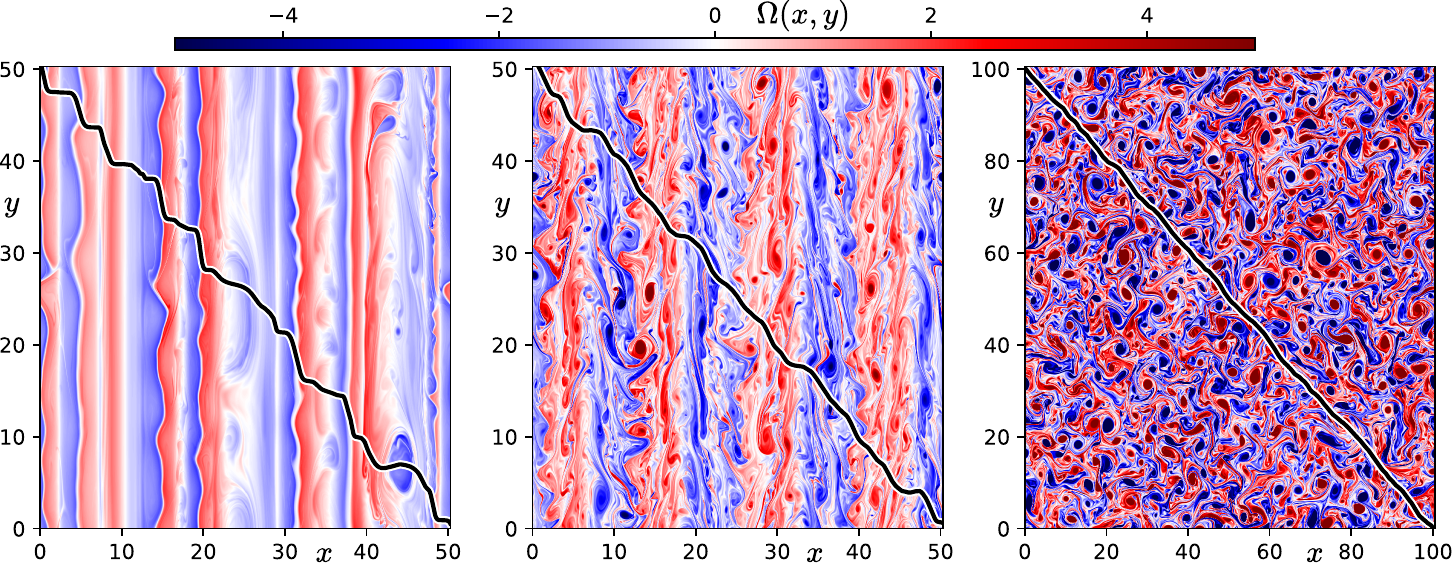}
\par\end{centering}
\caption{\label{fig:snaps}The snapshot of vorticity and the corresponding
PV staircase, for $C=10$, $C=1.0$ to $C=0.1$ from left to right,
for Hasegawa Wakatani turbulence. As we scan $C$ the transition scale
$k_{c}=C/\kappa$, changes also changing the zonostrophy parameter
(i.e. here effectively $R\propto k_{\text{peak}}/k_{c}$). We go from
clean staircases for large $C$ to a corrugated dirt hill to an almost
unmodified PV gradient. Comparing the relative overall levels of the
zonal to non-zonal spectra shown in Figure \ref{fig:spectra}, around
the peak of the nonzonal component shows this trend clearly. }
\end{figure*}
\begin{equation}
E\left(k\right)\approx\frac{\omega_{k}^{2}U\left(\gamma_{k}+\nu k^{2}\right)}{\left(\omega_{k}^{2}k^{2}+U\left(\gamma_{k}+\nu k^{2}\right)k^{3}\right)}\;\text{.}\label{eq:genform}
\end{equation}
The spectrum of Eqn. (\ref{eq:genform}), appears to be a very good
match to the results of numerical simulations as can be seen in Figure
\ref{fig:spectra}. It seems that this form of the wave-number spectrum
for plasma turbulence with zonal flows is rather generic in the adiabatic
(or high zonostrophy) limit. In the complete opposite limit of $k_{c}\ll1$,
we get $2D$ Navier-Stokes dynamics with injection around $k_{y}\approx1$,
and a forward cascade of enstrophy, but with no discernible inverse
cascade of energy (a key difference to linear inversion, or forced
$\beta$-plane\citep{scott:23}), where the spectrum in the inertial
range has the form $E\left(k\right)\propto\epsilon_{W}^{2/3}k^{-3}$,
where the coefficient $\epsilon_{W}^{2/3}$ can be computed directly
from enstrophy dissipation $\epsilon_{W}\equiv\sum_{k}\nu k^{4}E\left(k\right)$,
which comes mainly from small scales since $E\left(k\right)$ drops
slower than $k^{-4}$. Note that the faster than $k^{-3}$ drop of
the hydrodynamic spectra in Fig. \ref{fig:spectra} can be attributed
to interactions with large scales, a spectrum of the form $E\left(k\right)\sim k^{-4}/U$,
which may be related to the Saffmann spectrum as a consequence of
sharp vorticity gradients\citep{kuznetsov:07}, seems to be a good
fit as well, especially towards the dissipative range. 

In between the two limits is $C\approx1$, which is probably the more
challanging, and interesting case, in which we can still argue for
the standard forward enstrophy cascade picture for small scales (or
$k^{-4}$ due to sharp vorticity gradients), but since now we have
the injection scale and the transition scale $k_{c}$ basically on
top of each-other with zonal flows forming at slightly larger scales,
we can have a steady state with large scale zonal flows, more or less
governed by adiabatic limit of the equations and the small scale eddies
that are governed by 2D hydrodynamics. The stationarity of this state
with very strong zonal flows is extremely intriguing. A local linear
analysis suggests that the gradient of vorticity provides a feedback
loop which reduces the linear growth rate for negative vorticity gradient
by flattening the PV profile, and introduces a hypodiffusion like
term for positive vorticity gradient, providing a large scale energy
sink in the troughs of the zonal velocity profile, which allows the
underlying 2D hydrodynamic turbulence to reach a steady state, which
explains the steadyness of the fluctuations. The stationarity of large
scale zonal flows suggests that the 2D turbulence evolves towards
exact solutions of the 2D Euler equations, such as Kida vortices\citep{kida:81},
since those can survive in sheared flows without any net momentum
transfer, or dipole vortices\citep{meleshko:94}, which are known
to be abundant in similar structure dominated turbulent states\citep{gurcan:23}
with deplated nonlinearity due to selective decay\citep{sipp:00,pushkarev:14}
relevant also in plasma edge\citep{militello:17}. In other words,
in order to have stationary zonal flows with eddies, the eddies must
minimize their Reynolds stress on large scale flows.

\paragraph*{The Zonostrophy parameter}

A key parameter that defines the strength of the zonal flows to small
scale eddies, is the so called zonostrophy parameter $R_{\beta}$.
In the context of the usual (say forced) $\beta$ plane turbulence,
$R_{\beta}$ is defined as the ratio of the Rhines scale $L_{R}$
to the transition scale $\ell_{\beta}$\citep{scott:12}, which denotes
the scale of transition from 2D inverse cascade at small scales to
Rossby wave turbulence at larger scales. Basically if we have a large
range of scales between the $\ell_{\beta}$ and the $L_{R}$, the
zonal flows dominate over eddies, whereas if the two scales are close
together it is the other way around.

In the generalized beta plane perspective of the plasma turbulence
with internal drive, the zonostrophy parameter can be defined as $R_{\beta}\approx k_{\beta}/k_{\text{peak}}$
or $k_{c}/k_{\text{peak}}$, using $k_{c}$ as a proxy for $k_{\beta}$.
Snapshots of the vorticity field and the resulting PV staircase in
Figure \ref{fig:snaps} shows clearly that as we change $k_{c}$ by
two orders of magnitude while keeping $k_{\text{peak}}$ mostly unchanged,
the zonostrophy parameter and the resulting ratio of zonal vs. nonzonal
energy around the peak region changes substantially. Note that since
$k_{\text{peak}}$ and $k_{\beta}$ are roughly the scales at which
the linear growth and linear frequency are balanced by the nonlinear
decorrelation time respectively (see table \ref{tab:scales}), if
$\gamma_{k}\approx\omega_{k}$ (i.e. far from marginal) around the
most unstable mode, the two scales are basically on top of one another,
which makes the zonostrophy parameter small. In contrast if $\gamma_{k}\ll\omega_{k}$
for the most unstable mode, the two scales are well separated, which
makes $R_{\beta}$ rather large. This explains the ubiquitous observation
of abundance of zonal flows near marginality (i.e. $\gamma_{k}\ll\omega_{k}$),
and eddies far from it (i.e. $\gamma_{k}\gtrsim\omega_{k}$) in plasma
turbulence. The time scale comparison also explains the relation to
Kubo number, which is the ratio of the turbulence decorrelation time
(i.e. either $\gamma^{-1}$ or $\tau_{n\ell}$ ) to the nonlinear
time.

\paragraph*{Results and Conclusions}

Plasma turbulence is interpreted as a $\beta$-plane model with nonlinear
inversion, giving the original dispersion relation when fully linearized,
or a $\beta$-plane model with a linear growth rate, when only the
inversion is linearized, or the fully self-consistent nonlinear problem
when the inversion is fully nonlinear. It was observed that linear
inversion while giving a simple $\beta$ plane model with growth rate
from a linear dispersion relation, lacks the key saturation mechanism
\emph{involving the backreaction of the flow on the linear growth}.
 It was noted that $k_{c}=C/\kappa$, the scale at which the linear
system switches from adiabatic to hydrodynamic, plays the role of
$k_{\beta}$ for this system. In the adiabatic limit, a weak wave
turbulence spectrum in the form $E\left(k\right)\propto\omega_{k}^{2}k^{-3}$is
shown to be combined with the peak in the form of Eqn. (\ref{eq:genform}),
which matches the observed spectra from direct numerical simulations
for $k<k_{c}$. In contrast for $k>k_{c}$ it was observed that a
Kraichnan-Komogorov spectrum of the form $\varepsilon^{2/3}k^{-3}$
was observed in the ``inertial'' range. It was argued that the ratio
$k_{\beta}/k_{\text{peak}}$ can be defined as the zonostrophy parameter
$R_{\beta}$, which explains the appearance of zonal flows near marginality
and eddies away from it, since $R_{\beta}$ increases as we approach
marginality. 
\begin{acknowledgments}
This work was granted access to the Jean Zay supercomputer of IDRIS
under the allocation AD010514291 by GENCI. The author would like to
thank the Isaac Newton Institute for Mathematical Sciences, Cambridge,
for support and hospitality during the programme \textquotedblleft Anti-diffusive
dynamics: from sub-cellular to astrophysical scales\textquotedblright{}
where work on this paper was undertaken. This work was supported by
EPSRC grant no EP/R014604/1.
\end{acknowledgments}

\end{document}